# Donor-acceptor recombination in diamond

Konstantin Iakoubovskii[a,*], Andrey Katrusha[b], Weihua Peng[b], Jianguo Peng[b]

[a] *Central European Institute of Technology, Brno University of Technology, Purkyňova 123, 61200 Brno, Czech Republic*
[b] *Jilin Diamond Technology Research & Development, Jianan Road 177, Luyuan District, Changchun, China*

**Abstract**
We re-examine several optical systems that have previously been attributed to donor–acceptor (D-A) recombination in diamond, including the broad 2.8 eV and 2.3 eV bands, and the sharp-lined series in the ranges 2.0-2.7 eV and 4.0-5.0 eV. In all cases, internal transitions at one or multiple defect centers provide a more consistent explanation of the experimental observations than the D-A model. The 2.3 eV case is marginal, where the key argument against the D-A recombination is the improbably low probability (~$10^{-100}$) of tunneling between remote donors and acceptors in diamonds with very low concentrations of impurities (~0.1 ppm). A more realistic model for the 2.3 eV band and the associated phosphorescence and thermoluminescence phenomena is trapping of electrons and holes at nitrogen and boron defects, respectively, followed by recombination at a deep center, which is tentatively assigned to the nearest neighbor B-N pair.

*Corresponding author: iakoubovskii@vut.cz

## 1. Introduction

Donor-acceptor (D-A) recombination is one of the most popular models for interpretation of luminescence in semiconductors. In this model, an electron bound to a donor recombines with a hole localized at an acceptor, producing signals whose energy $h\nu$ depends on the bandgap $E_g$, donor and acceptor ionization energies $E_D$ and $E_A$, and the Coulomb interaction between the ionized defects [1]:

$$h\nu = E_g - E_D - E_A + \frac{e^2}{4\pi\epsilon_0\epsilon r} - SE_{ph}. \qquad (1)$$

Here $e$ is electron charge, $\varepsilon_0$ and $\varepsilon$ are static dielectric constants of the vacuum and the semiconductor, respectively, and $r$ is the D-A separation. The last term accounts for emission of phonons, which is required if the ground states of donor and acceptor are shifted in the configuration space. This term can be simplified as a product of the average phonon energy $E_{ph}$ and the number of phonons emitted per transition $S$, which is also called the Huang-Rhys factor [1].

Via the Coulomb term, D-A recombination provides information on impurity distribution, compensation mechanisms, and defect interactions. The D-A model has been successfully applied to a wide range of semiconductors including Si [2], GaP [3,4], SiC [4], GaAs [5,6], CdS [7], ZnSe [8], ZnO [9,10], and GaN [9,11].

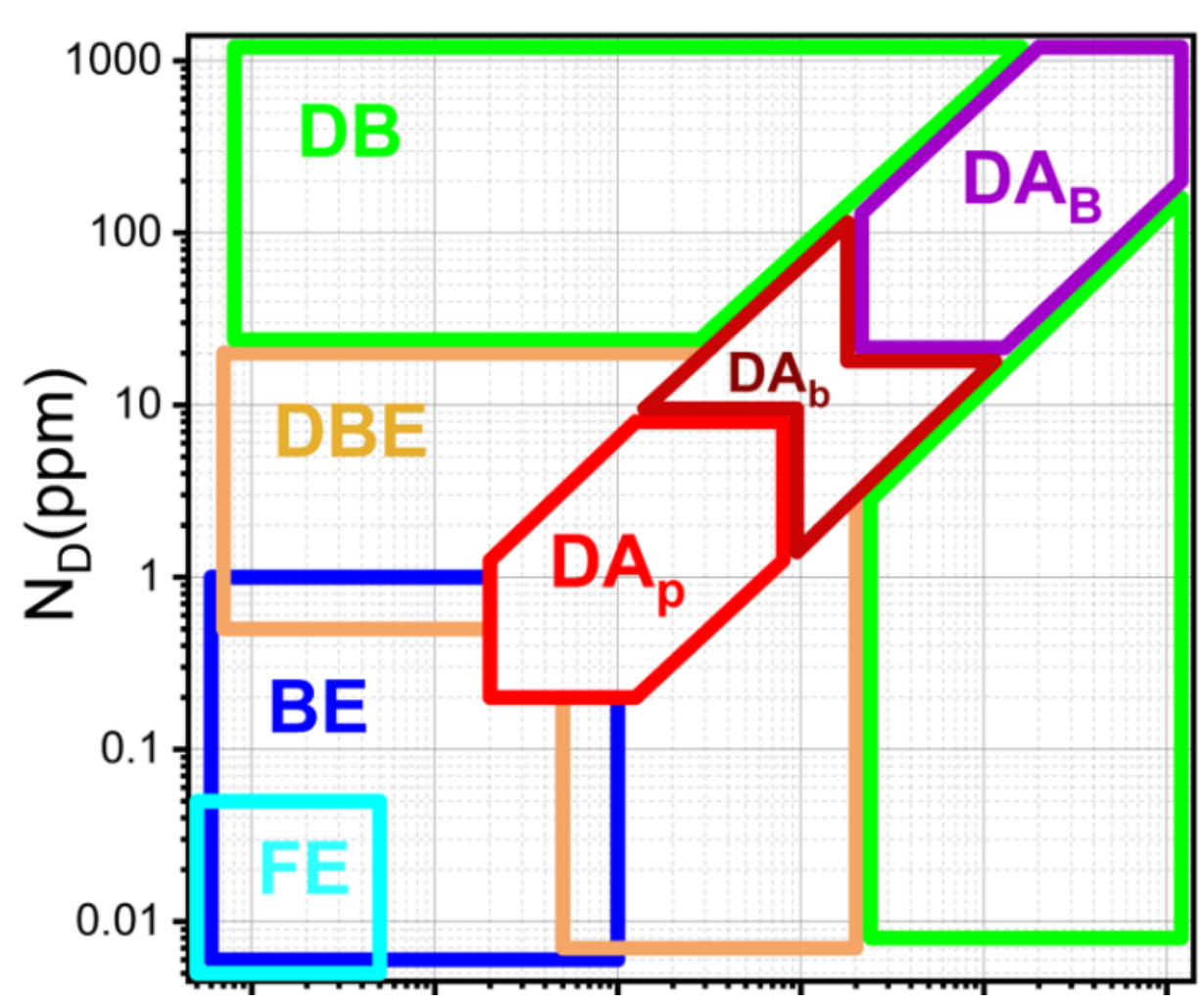

*Fig. 1. Schematic assignment of luminescence bands in silicon, an exemplary host of donor-acceptor pairs[2]. Legends: FE – free exciton, BE – exciton bound to a donor or acceptor, DBE – exciton bound to another defect, DB – transitions involving degenerate bands that form at high concentrations of donors.*

The physics of D-A recombination is particularly well established in classical semiconductors such as silicon. In compensated Si containing phosphorus donors and boron acceptors, the luminescence spectrum evolves systematically with impurity concentration (Fig. 1). At high concentrations, the spectrum is dominated by closely spaced D-A pairs and exhibits broad bands shifted to higher energies owing to the Coulomb interaction. As the concentration decreases, the average D-A separation increases, the contribution of distant pairs becomes more important, and the spectrum shifts toward the limiting transition energy corresponding to infinitely separated donors and acceptors. In material with less than 10 parts per million (ppm) of impurities, discrete spectral lines associated with specific D-A shells can be resolved and directly related to the crystallographic positions of the impurities; these lines disappear at concentrations below ca. 0.2 ppm. The transition energies, spectral evolution, and recombination kinetics are then described successfully by a relatively simple model in which the donor and acceptor behave as localized point charges embedded in a dielectric medium [2].

The success of this approach relies on several assumptions that are generally satisfied in conventional semiconductors: donor and acceptor states are shallow and hydrogenic, with wavefunctions extending over several lattice constants. The Coulomb interaction between the ionized defects is therefore the dominant contribution to the variation of transition energies, while the phonon term in Eq. 1 is comparatively small. As a result, the transition energy can be interpreted in terms of D-A separation alone, and the relationship between optical spectra and defect geometry is relatively straightforward [2].

Diamond differs fundamentally from these classical systems. Owing to its exceptionally strong covalent bonding, large bandgap of 5.47 eV, and high Debye temperature and phonon energy, most impurity states in diamond are deep and strongly localized. The two most common electrically active impurities, substitutional nitrogen and substitutional boron, represent a deep donor and an acceptor, respectively, whose wavefunctions are confined to only a few atomic spacings. Such strong localization has several important consequences for donor–acceptor recombination. First, the overlap of donor and acceptor wavefunctions decreases rapidly with separation, making radiative transitions between widely separated impurities intrinsically inefficient. Second, the localized nature of the electronic states enhances coupling to lattice distortions, leading to substantial configurational relaxation. Unlike many conventional semiconductors, where relaxation energies are often small compared with Coulomb energies (see last two terms in Eq. 1), optical transitions involving deep defects in diamond may be accompanied by relaxation energies exceeding 1 eV [12]. Under such conditions, the transition energy can no longer be determined solely by donor and acceptor ionization energies and their Coulomb interaction, and the direct correspondence between transition energy and donor–acceptor separation becomes less obvious [1].

In this work, we re-examine the D-A recombination model in diamond on examples of several systems that were considered previously.

## 2. Experimental details

HPHT diamond crystals were synthesized at 6.5–7.0 GPa using a cubic press and temperature gradient method. Temperature was varied between 1450 and 1750 ºC. We used a Co-Fe, Co-Ni, or Fe solvent catalyst and a graphite carbon source, with purity >99% for all materials. Nitrogen content in the samples was varied by adjusting the amount of an aluminum-based nitrogen getter. Intentional boron doping was achieved by introducing amorphous boron to the growth mixture. All seeds were HPHT diamonds; they were selected to have prominent (111) facets, and these facets were aligned to the direction of thermal gradient in order to create crystals with predominant (111) sectors (see inset in Fig. 3). This measure would reduce the sample inhomogeneities due to dissimilar incorporation efficiencies of nitrogen and boron into different growth sectors. The as-grown samples were thoroughly cleaned in mixtures of sulfuric, nitric and hydrochloric acids, and their front facets were polished for optical absorption studies.

Polycrystalline diamond films were grown by chemical vapor deposition (CVD) on silicon substrate using an ASTeX PDS-17 microwave plasma reactor at the Institute for Materials Research in Belgium. After growth, the substrates were removed by chemical etching in $HNO_3$:HF acid mixture.

Crystalline quality of diamonds was assessed by Raman spectroscopy using a WITec alpha 300R confocal microscopy setup equipped with 633 nm (HeNe), 355 and 532 nm (Nd:YAG) lasers. The Raman setup was also used for some photoluminescence measurements.

Concentrations of nitrogen and boron were measured by Fourier-transform infrared (FTIR) absorption with a Vertex70v spectrometer from Bruker. Measurements were performed in vacuum to avoid atmospheric instabilities. The presence of other impurities was assessed by energy-dispersive X-ray (EDS) and time-of-flight secondary ion mass (ToF-SIMS) spectroscopies inside an Amber X2 scanning electron microscope (Tescan). Detection limits were approximately 0.1, 1000, and 0.1 ppm for FTIR, EDS and SIMS, respectively.

Phosphorescence and thermoluminescence were excited with a monochromatized deuterium lamp (220 nm, 10 nm bandwidth) or with a CuKα X-ray source (Rigaku MiniFlex, 40 kV, 15 mA, 0.15406 nm) and the spectra were recorded with a USB2000+ spectrometer from Ocean Optics. In thermoluminescence runs, we first irradiated the sample at 77 K, waited for phosphorescence to decay, and then heated the sample up at a rate of 1 K/s, while measuring intensity at 2.5 eV (500 nm) that was taken from 200-890 nm spectra acquired with the USB2000+. Jasco V-770 spectrometer was employed for UV-visible optical absorption and luminescence excitation measurements.

Time-resolved luminescence was excited with a nitrogen laser, which provided 3 ns 0.3 mJ pulses at 3.68 eV (337 nm), and detected with a photomultiplier on a Tektronix oscilloscope. Cathodoluminescence (CL) was excited by a 40-keV electron beam in the setup built at Kings College London.

## 3. Analysis of potential D-A bands

### 3.1. The 2.8 eV band

Most diamonds exhibit broad luminescence bands centered at 2.8 and/or 2.3 eV at 77 K, which were originally named the A bands. In 1965, Dean proposed that both these bands originate from recombination between two sets of donor-acceptor pairs: closely spaced and distant, respectively [13].

Figure 2 shows low-temperature photoluminescence (PL) and PL excitation (PLE) spectra of the 2.8 eV band in HPHT diamond with boron concentration [B] = 0.2 ppm, and concentrations of nitrogen [N] and other impurities below the detection limit (0.1 ppm for that sample). Both PL and PLE spectra are featureless and are almost mirror images of each other. Two important observations should be added to that. First, the lifetime of the luminescence band was measured as 10 ns, in a good agreement with a previous report [14]. Second, no 2.3 eV band could be detected in this sample under any excitation energy in the range 2.5-5.5 eV. The measured lifetime likely approximates the intrinsic radiative lifetime of the transition. Indeed, if it was due to some competing non-radiative process, then the luminescence would probably be undetectable in excitation measurements, where the light intensity was < 5 mW/cm$^2$.

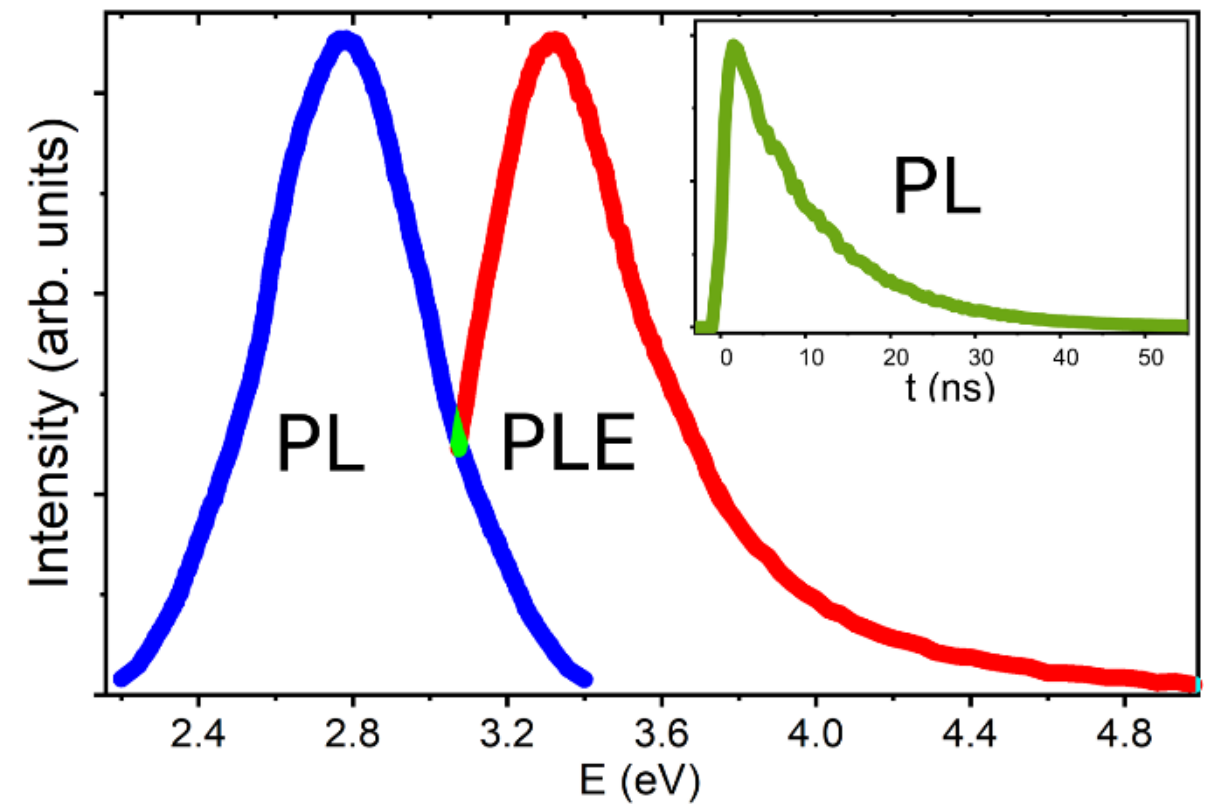

*Fig. 2. Photoluminescence (PL, 3.6 eV excitation) and PL excitation (PLE, 2.8 eV emission) spectra (80K ) of the 2.8 eV band in HPHT diamond with [B] = 0.2 ppm and [N] < 0.1 ppm. Inset shows PL relaxation under nitrogen laser excitation (3.68 eV, 3 ns).*

These observations seriously question the D-A assignment for the 2.8 eV band. Indeed, its lifetime is much too short for distant D-A pairs. However, if this band corresponds to close pairs, then it is unclear why distant pairs are not seen in the same sample, especially considering its low impurity content.

This conclusion agrees with the literature, especially with cathodoluminescence measurements that spatially correlate dislocations with the 2.8 eV but not the 2.3 eV band [15-19]. Given the symmetric line shapes of the luminescence and its excitation spectra, we speculate that the 2.8 eV band originates from a vibronic system, an internal transition at a deep-level optical center pinned at dislocations, in which the zero-phonon line is broadened by the dislocation-related strain field.

### 3.2. The broad 2.3 eV band

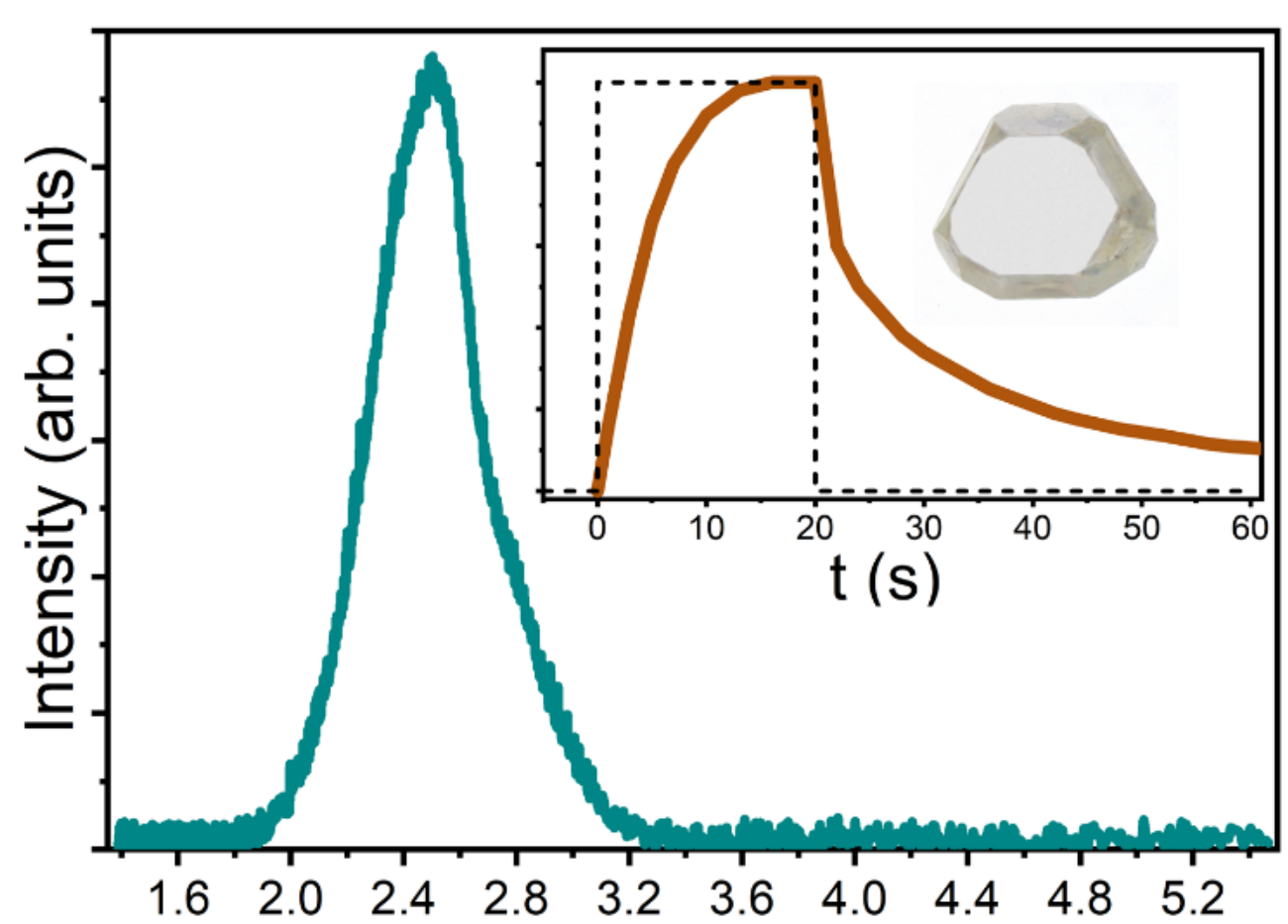

*Fig. 3. Room-temperature X-ray excited luminescence spectrum from an HPHT diamond with impurity content <0.1 ppm. Inset shows the luminescence kinetics after applying a square X-ray pulse and a photograph of the sample. The dashed line represents the X-ray intensity*

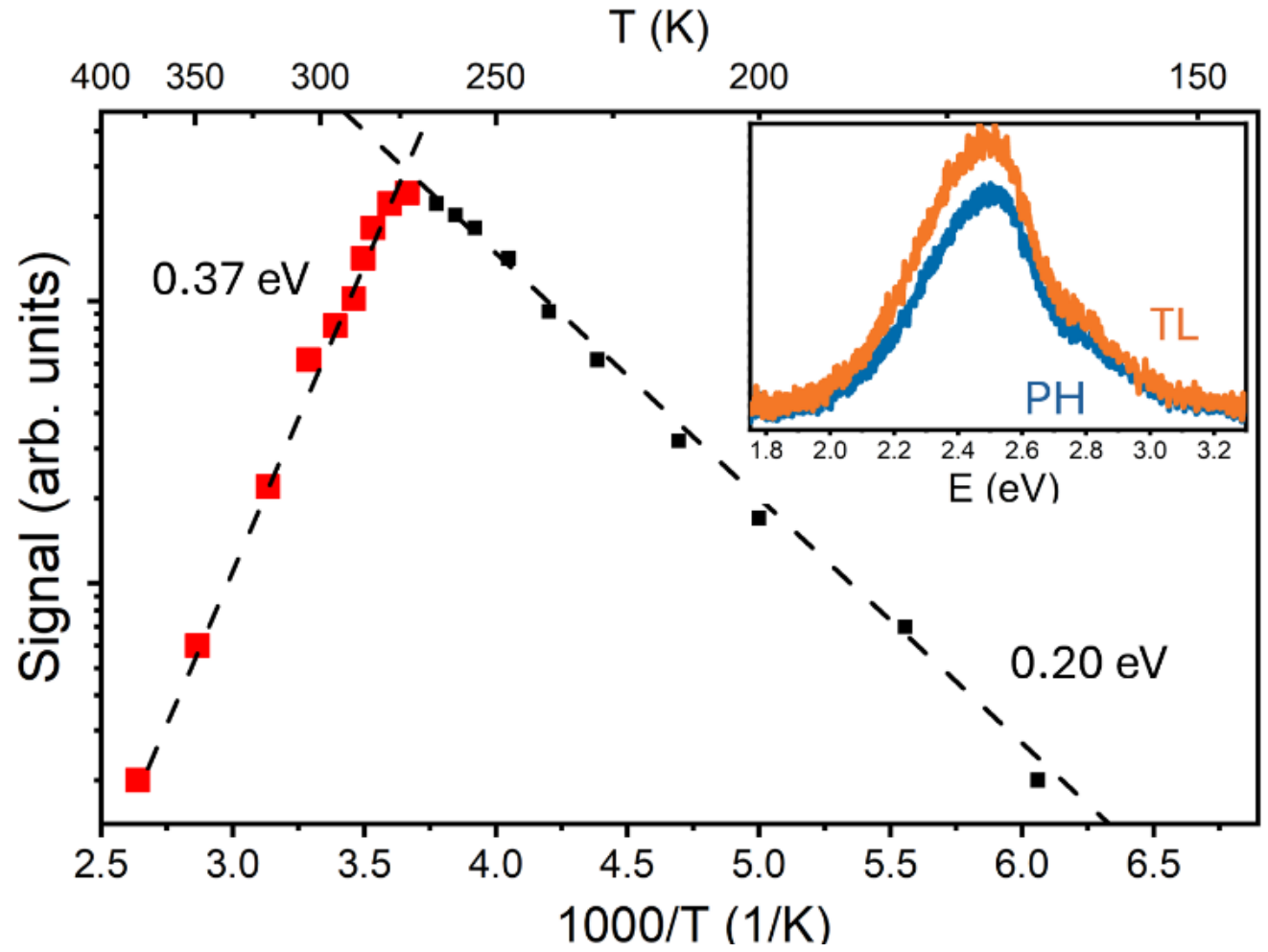

*Fig. 4. Temperature dependence of phosphorescence from an HPHT diamond with impurity content <0.1 ppm. Inset shows room-temperature thermoluminescence (TL) and phosphorescence (PH, 220 nm excitation) spectra.*

While the Dean's assignment of the blue 2.8 eV band to D-A recombination was frequently criticized [14,20-22], the D-A nature of the 2.3 eV band had been widely accepted [14,20-22], and only strengthened with the recent detailed studies by the groups of M. Newton [1,23] and A. Shen [24,25] that unambiguously linked the cathodoluminescence, photoluminescence, phosphorescence and thermoluminescence properties of the 2.3 eV band with the nitrogen donor and boron acceptor. Especially informative is the study [1], which was however based on two parts of a single HPHT sample. In this work we reproduced those results on two dozens of HPHT samples with different concentrations of nitrogen and boron. We confirm the association of the band with these impurities, but not its D-A recombination nature.

The following discussion focuses on HPHT samples with impurity content below 0.1 ppm, which is somewhat lower than that of Newton *et al.* (0.145 ppm [1]). These samples exhibited a Raman linewidth of 1.5 $cm^{-1}$ and no luminescence under excitation with wavelength above ca. 230 nm (see Figs. S1 and S2 in the Supplemental Information). Upon room-temperature excitation with bandgap light or X-rays such samples exhibited a broad band centered at ca. 2.5 eV, as shown in Fig. 3. Cooling the sample down to 77 K resulted in its shift to ca. 2.3 eV, in agreement with previous reports [1, 26], while no fine structure could be detected. The same band was detected in thermoluminescence. As shown in Fig. 4, this broad band featured seconds-long kinetics and unusual temperature dependence that peaked upon heating at ca. 280 K and had activation energies of 0.20 eV below and 0.37 eV above the peak.

These results agree very well with previous reports [1, 23-25] and strongly suggest that the 2.3 eV emission is linked with levels at $E_C$-0.2 eV and $E_V$+0.37 eV, which were previously assigned to the $N_S^-$ and $B_S^0$ centers [1]. After bandgap excitation at low temperatures, these defects trap electrons and holes, respectively, and virtually no luminescence is observed until heating releases electrons from the $E_C$-0.2 eV level into the conduction band. This relatively high activation energy explains the slow kinetics of emission (see inset in Fig. 3), and does not reflect the recombination lifetime. Subsequent thermal excitation of holes into the valence band quenches the luminescence.

This scenario is fully consistent with the D-A model for the 2.3 eV band, where the donor is nitrogen and acceptor is boron (see Fig. 5). Due to statistical fluctuations of the Fermi level in a wide-bandgap insulator [27], at least four charge states are present: $N^+$, $N^0$, $B^-$ and $B^0$. After bandgap excitation at low temperature, electron-hole pairs are created. Holes are captured by $B^-$ creating $B^0$ and electrons are trapped by $N^0$ forming $N^-$ ($E_C$–0.2 eV level). Upon heating up to 280 K (see Fig. 4), electrons are released from $N^-$, captured by $N^+$ thereby creating $N^0$ donors, and tunnel to $B^0$ acceptors resulting in D-A emission. This emission is quenched upon further heating above 280 K, as it ionizes the acceptors.

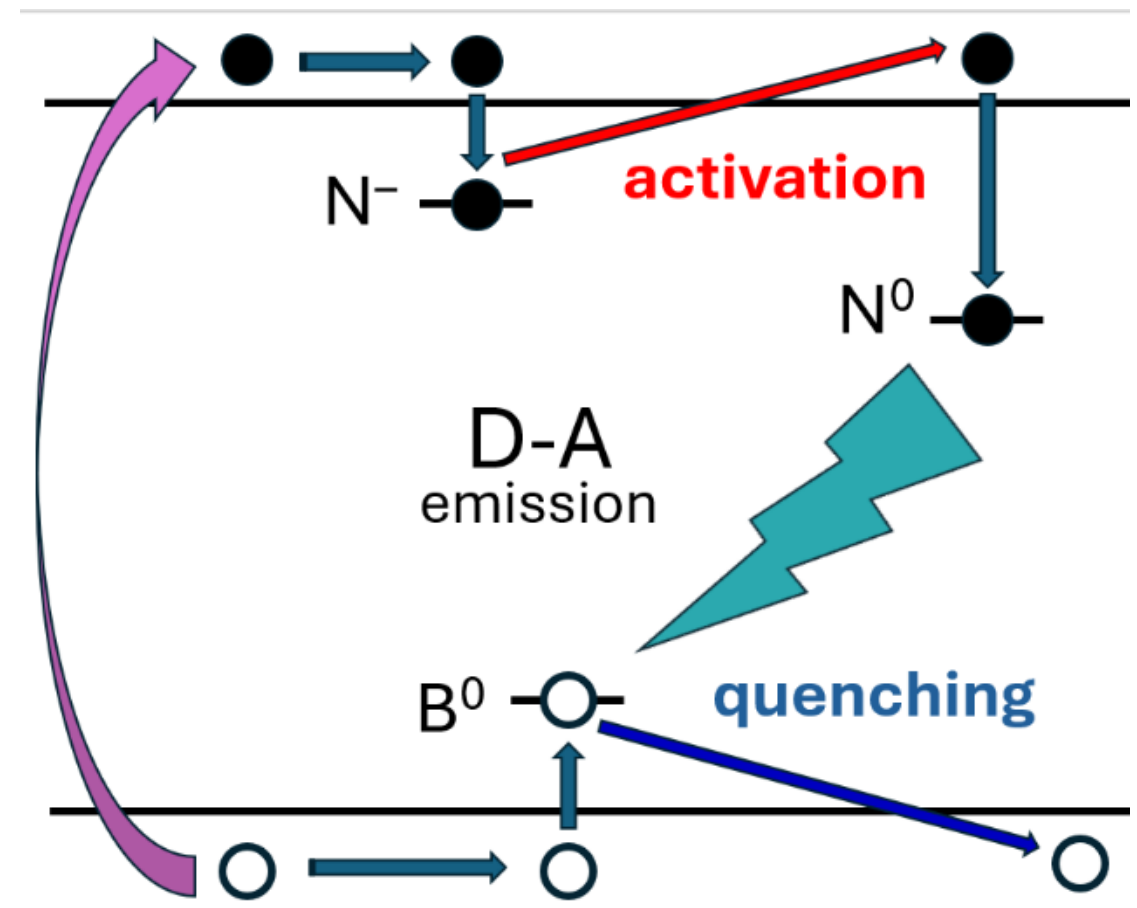

*Fig. 5. Schematic of the D-A model for the 2.3 eV band*

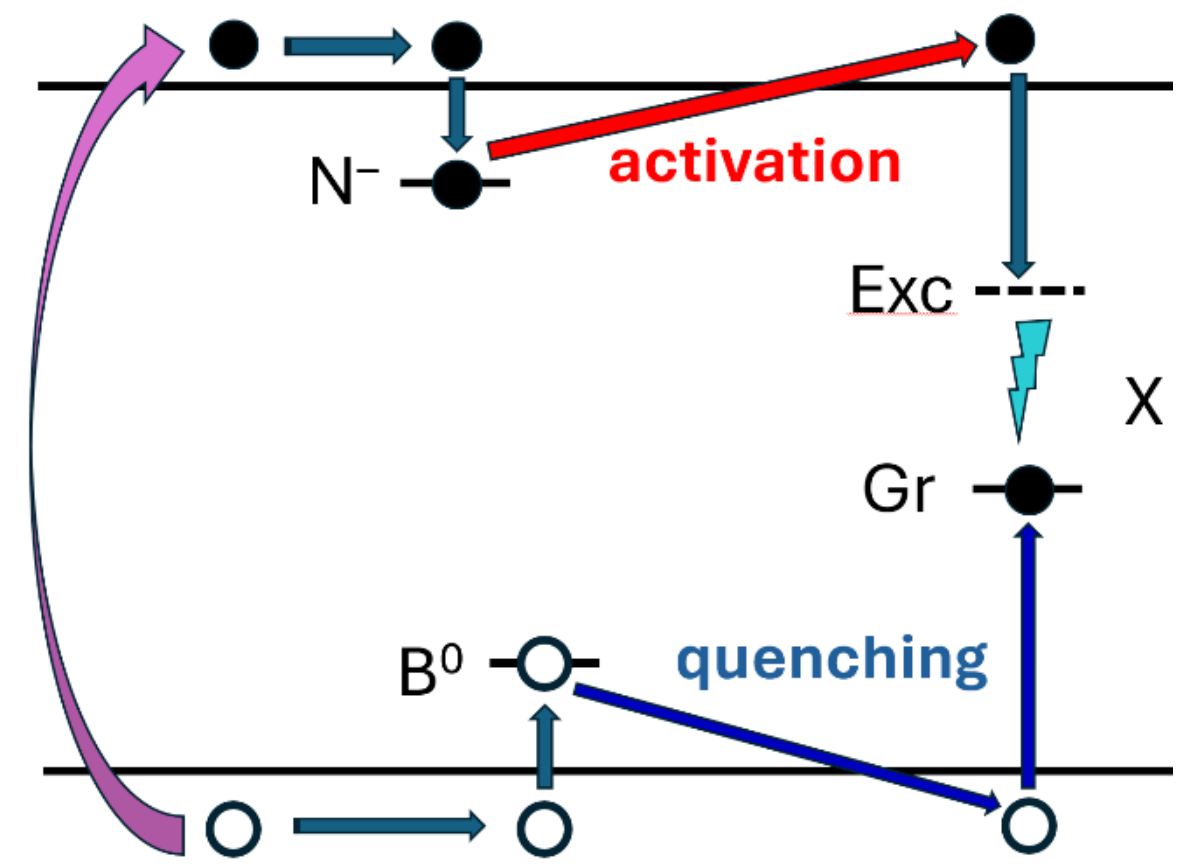

*Fig. 6. Alternative model for the 2.3 eV band, where all trapping processes remain the same as in the D-A model, and the emission originates not from D-A pairs, but from an internal transition from an excited state Exc to the ground state Gr at a defect X that we tentatively assign to the nearest-neighbor B-N pairs.*

While this D-A model naturally explains all the observed thermal dependences and photochromic changes in optical absorption and electron spin resonance signals [1], it fails to explain how D-A emission could be observed in diamonds with concentration of donors and acceptors as low as 0.1 ppm or even less, which was never reported for other semiconductors (see also Fig. 1). At such concentrations, the average distance between defects R exceeds ca. 38 nm. Meanwhile, the D-A tunneling probability Q is proportional to exp(-2R/*a*), where *a* is the Bohr radius for the shallower impurity (boron) [28]. This radius can be calculated as $a = e^2/(4\pi\varepsilon_0\varepsilon E_B) = 0.34$ nm [29], where $\varepsilon \sim 5.7$ is the static dielectric constant and $E_B = 0.37$ eV is the ionization energy of the boron acceptor. Thus, we arrive at the tunneling probability Q on the order of 0.1 for closest pairs and $10^{-100}$ for those separated by 38 nm. Such low transition rates are unobservable in a realistic experiment, and have not been reported in any semiconductor, to the best of our knowledge. The lowest concentrations for observation of D-A pairs in Si are at the level of 0.2 ppm, where *a* = 1.6 nm is much larger than in diamond, and hence the tunneling probability is much smaller, on the order of $10^{-16}$.

The above inconsistency can be easily resolved by assuming that the 2.3 eV emission originates not from D-A recombination, but from a localized transition at some low-density defect X (see Fig. 6). In this model, boron and nitrogen remain as trapping centers, and hence all thermal dependences and photochromic changes remain accounted for, yet there is no assumption of tunneling, and hence the model holds for a wide range of defect concentrations. This defect X captures electron re-emitted from $N^-$, and generates luminescence, which is then quenched upon capture of a hole re-emitted by $B^0$.

The microscopic model of defect X can be tentatively assigned to the nearest-neighbor B-N pair. This assignment would naturally explain the simultaneous presence of defect X, boron and nitrogen. Theory predicts that the formation of such nearest-neighbor B-N pairs is more favorable as compared to distant pairs [30-32].

### 3.3. The structured 2.3 eV band

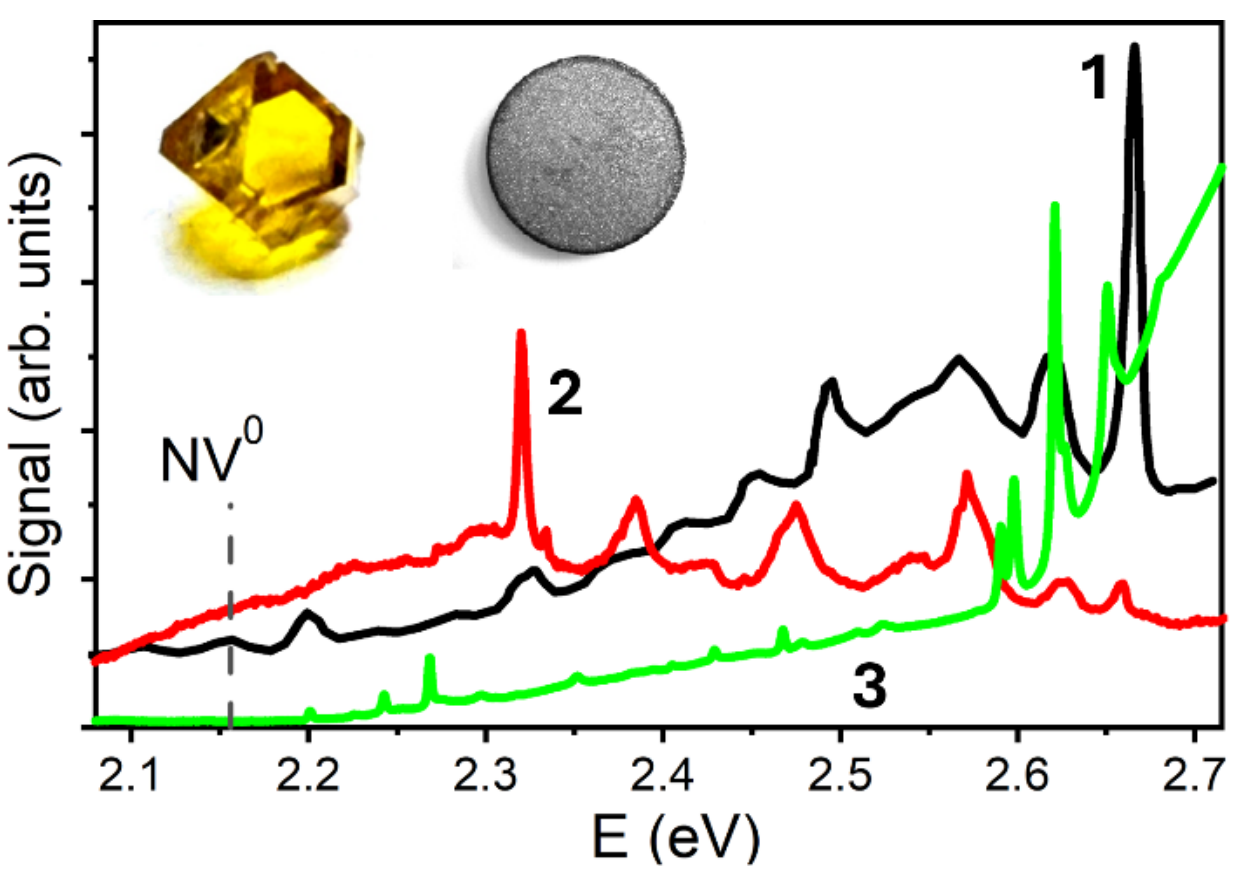

*Fig. 7. CL spectra from CVD films reported by Dischler et al. [33] (curve 1) and studied in this work (2). Curve 3 shows absorption spectrum from an HPHT diamond similar to that referenced by Dischler et al. All spectra were recorded at 77 K. Inset shows photographs of the studied samples.*

In 1994–1995, Dischler *et al.* reported a series of sharp luminescence and absorption lines in the range 2.1–2.7 eV that fitted surprisingly well into the D-A model (Eq. 1) [33-35], with nitrogen and boron being the donor and acceptor, respectively. Here we reexamine these results and suggest that those lines are probably unrelated to each other.

Dischler *et al.* compared CL spectra from CVD diamond films, grown in a microwave plasma reactor, with absorption spectra from HPHT crystals. In this work, we partly reproduced their results on similar samples. Figure 7 compares CL spectra from CVD films taken from ref. [33] and CVD films measured in this work. We also add an absorption spectrum from an HPHT diamond similar to that used as a reference by Dischler *et al.* A broad agreement is seen between these spectra; however, detailed analysis of the studied samples strongly suggests that the reported series consist of unrelated peaks.

First, the reference HPHT diamonds were grown in Ni-containing solvent without a getter, and hence were rich both in nickel and nitrogen. Removal of Ni from the growth medium resulted in disappearance of the questioned lines, and hence they can be assigned to Ni. Their relative intensities strongly change upon post-growth annealing at temperatures above 1600 °C, strongly suggesting multiple Ni-related defects rather than D-A pairs as their origin [36].

Second, while the CVD diamond films studied by Dischler *et al.* were intentionally doped with nitrogen and boron, our films were grown in a reactor free from boron and nitrogen, and these impurities could not be detected in the resulting film by SIMS down to ppm level. This could be indirectly confirmed in Fig. 7 via the absence of the 2.156 eV peak originating from the neutral nitrogen-vacancy complex ($NV^0$) in our films. Our measurements on a series of samples suggest that the studied CL peaks are individual and do not form a series. For example, the peaks at 2.67, 2.59 and 2.49 eV mostly originate from the interstitial-related defect TR12' [37]. All these results strongly suggest that the peaks reported by Dischler *et al.* originate from several individual optical centers rather than from D-A recombination.

### 3.4. The 4.6 absorption band

Besides the nickel-related system mentioned in the previous section, numerous other optical absorption series are known in diamond that do resemble D-A systems. One of them, the 4.6 eV system, is worth mentioning because it universally appears in CVD and HPHT diamonds that contain single nitrogen, even at concentrations down to sub-ppm level. It consists of a broad band and a series of sharp peaks, with both components having trigonal symmetry [38,39]. Hence, the 4.6 eV system was assigned to the substitutional nitrogen donor [38,40,41], which invariably appears in all D-A models in diamond.

Figure 8 shows the 4.6 eV absorption band in two high-quality HPHT diamonds, as judged by the narrow width of the diamond Raman peak (<2 $cm^{-1}$). The top curves shows a series of sharp peaks that might imply a D-A series, yet the bottom curve suggests that the whole structure is merely a vibronic system with a zero-phonon line at 4.064 eV and a phonon replica of 60 meV.

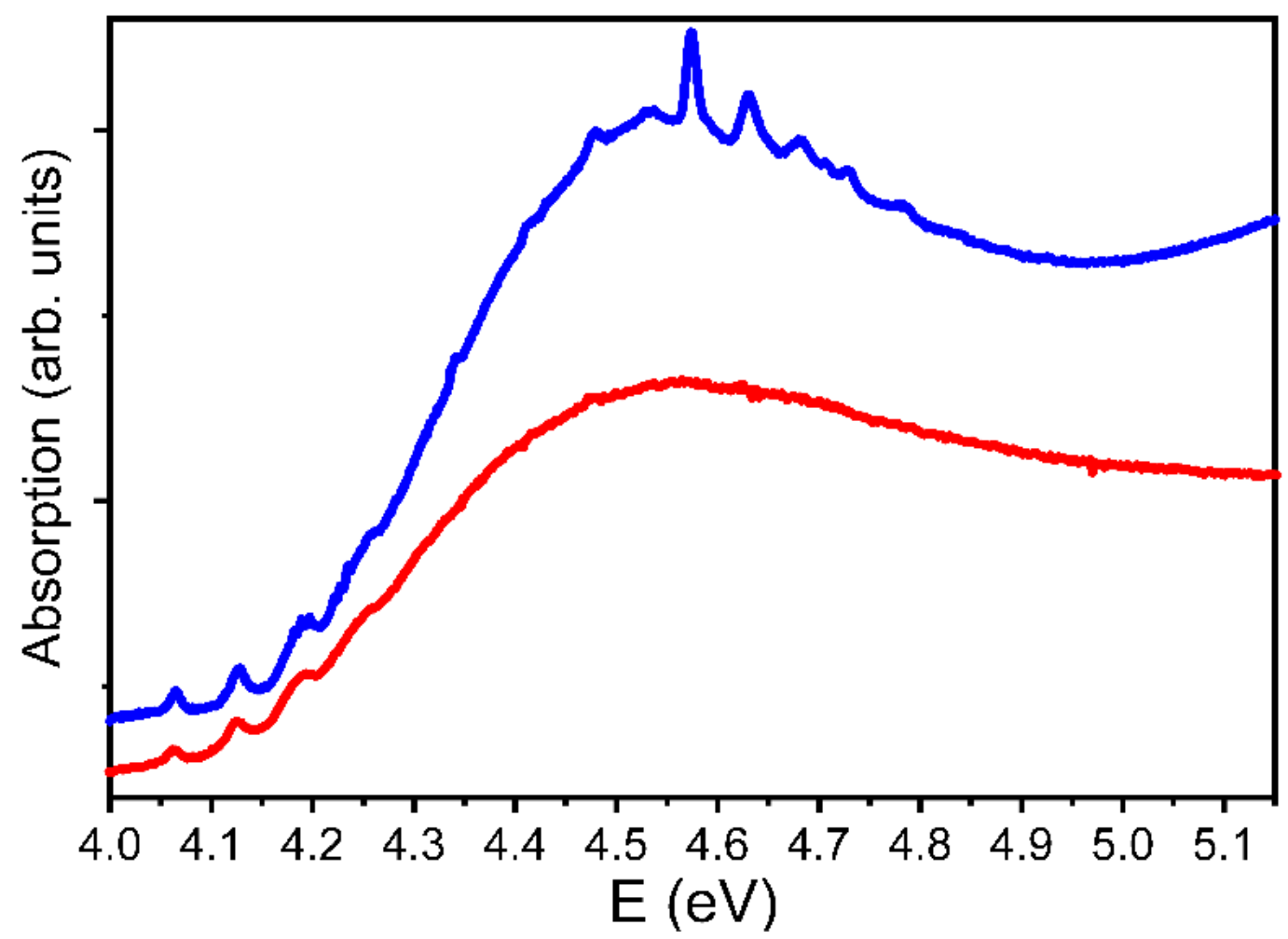

*Fig. 8. Optical absorption (77 K) of the 4.6 eV system in HPHT diamonds grown at different temperatures (ca. 1400 °C for the top and 1500 °C bottom curves).*

## 4. Conclusions

We have examined several optical systems that might originate from D-A recombination in diamond, including the broad 2.8 eV and 2.3 eV bands, and the sharp-lined series in the ranges 2.0-2.7 eV and 4.0-5.0 eV. In all cases, internal transitions at one or multiple defect centers explain experimental results better than the D-A model. The only marginal case is the 2.3 eV band, where the key argument against the D-A recombination is the improbably low probability of tunneling between remote donors and acceptors in diamonds with very low concentrations of impurities.

## Acknowledgments

CzechNanoLab project LM2023051 funded by MEYS CR is gratefully acknowledged for the financial support of the measurements at CEITEC Nano Research Infrastructure. K. I. is grateful to Prof. M. Nesládek for provision of CVD samples and Prof. A. Collins for assistance with CL measurements at KCL.

## Supplementary material

Supplementary data for this article can be found online at ...